\documentclass[runningheads]{svmult}
\usepackage{makeidx}   
\usepackage{graphicx}  
\usepackage{subeqnar}  
\usepackage{multicol}  
\usepackage{cropmark} 
\usepackage{physprbb}  
\makeindex             
\def\ltsima{$\; \buildrel < \over \sim \;$}
\def\simlt{\lower.5ex\hbox{\ltsima}} 
\def\gtsima{$\; \buildrel > \over \sim \;$}
\def\simgt{\lower.5ex\hbox{\gtsima}} 
                                                          
\begin{document}
\title*{Optical Observations of Afterglows}
\toctitle{Focusing of a Parallel Beam to Form a Point
\protect\newline in the Particle Deflection Plane}
%
%
\titlerunning{Optical Observations of Afterglows}
%
\author{Elena Pian 
}
\authorrunning{E. Pian}
%
%
\institute{INAF, Astronomical Observatory of Trieste, Via G.B. Tiepolo 11, I-34131
Trieste,
Italy
}

\maketitle              

\begin{abstract}

Optical, infrared and ultraviolet observations of GRB fields have
allowed detection of counterparts and host galaxies of the high energy
transients, thus crucially contributing to our present knowledge of
the GRB phenomenon.  Measurements of afterglow variable emission,
polarized light, redshifted absorption and emission line spectra, as
well as host galaxies brightnesses and colors have clarified many
fundamental issues related to the radiation mechanisms and
environments of GRBs, setting the background towards disclosing the
nature of their progenitors.

\end{abstract}

\section{Introduction}

GRB counterparts are multiwavelength emitters, unlike supernovae,
which emit most of their power at the optical frequencies.  However,
observations in the optical band have had the biggest impact in the
study of the GRB phenomenon, by sampling the time histories of the GRB
counterparts from a few seconds up to years after the explosion, by
establishing the nature of the afterglow emission (synchrotron
radiation), and by unambiguously assessing the extragalactic origin of
GRBs.  GRB optical counterparts can be initially extremely bright. 
However, the delayed emission, even though it is long lived with
respect to the prompt event, still fades quite rapidly, and the host
galaxies of GRBs are very small and faint, because of their
cosmological distances.  Therefore, telescopes of all aperture sizes
have been involved in the investigation of the optical afterglows. The
smaller telescopes are more flexible for an efficient search and the
larger ones (including HST) play a leading role in early spectroscopy
and polarimetry, as well as in the photometric monitoring at late
epochs, and in the study of the host galaxies. The most important
observations of optical afterglows and their host galaxies, which have
led to the current understanding of the physics of these sources, are
discussed. Previous reviews on this subject include
\cite{vkw00,klose00,galama00,kbb+00,pab+00,dql00,ajct01,sfc+01,dkb+01,sh01}.
Although the present review focuses on the optical observations,
information provided by the scanty infrared and ultraviolet data is
also included, considering the proximity of these bands to the
optical. The chapter is organized as follows: in Section 2 the steps
which led to the first detection of optical afterglow emission are
summarized and the observational problems related to the afterglow
investigation are described; in Section 3 the temporal behavior of the
optical prompt and afterglow emission is reviewed; in Section 4 the
observed spectral continuum at infrared, optical and ultraviolet
wavelengths is compared to the standard afterglow model based on the
propagation of an external shock, and the characteristics of the
afterglow absorption spectra are illustrated.  Section 5 is devoted to
a discussion of those aspects of GRB host galaxies which seem to be
relevant to the afterglow investigation. Future perspectives in the
exploration of GRB optical counterparts are sketched in Section 6.

\section{Optical searches of GRB error boxes}

The search for optical counterparts of GRBs started soon after GRB discovery in
the belief that they held the key toward the physical origin of the GRB
phenomenon. The first searches in the 1970's were based on inspection, on archival
plates or on photographic images of the GRB field, of the large error boxes
yielded by the early high energy missions (see \cite{vkw00} for a review of these
pioneering attempts). The reasons of their failure, as we now understand, reside
in the inadequacy of the methods.  Since GRB afterglows fade quickly, only timely
and sensitive surveys of accurate, rapidly disseminated GRB error boxes can be
effective in detecting optical counterparts.

The first identification of an optical afterglow was made possible on 28 February
1997 by the BeppoSAX satellite \cite{bbp+97}, whose Wide Field Cameras (WFC,
\cite{jmb+97}) determined the position of a GRB with a $3^{\prime}$ uncertainty
radius and disseminated it to the community within a few hours \cite{cfh+97}.
Optical observations made 21 hours and $\sim$1 week after the GRB allowed the
detection of a variable source in the GRB error area, which had been meanwhile
refined and reduced to $\sim$1 arcmin$^2$. Variability and positional coincidence
suggested association of the optical transient with the GRB \cite{vgg+97}. Five
years later, out of about 100 GRBs accurately and rapidly localized by BeppoSAX or
by other spacecrafts, and timely followed by optical telescopes (delays of a few
hours to a few days of the GRB explosion), $\sim$30 have a detected optical afterglow.  
These are reported in Table~1, together with those GRBs for which, despite the
lack of optical afterglow detection, the precise afterglow positioning at other
wavelengths has allowed the identification of a host galaxy (in these cases the
lower limit on the magnitude of the transient counterpart is reported, as
determined in the earliest search). Col. 2 of Table~1 reports the instrument, or
suite of instruments, which have localized the GRB:  ``BeppoSAX" means that the
event has triggered the Gamma-Ray Burst Monitor (GRBM) of the BeppoSAX satellite
\cite{fcd+97}, and has been localized with arcminute precision by the onboard WFC.  
In two cases the BeppoSAX WFC detected in the X-rays and localized an event which
triggered BATSE, but not the GRBM.  For almost all GRBs localized by the Rossi XTE
satellite, the accurate positioning was obtained with its All Sky Monitor, the
exception being GRB990506, which was localized precisely only after the RXTE
Proportional Counter Array (PCA) detected its X-ray afterglow \cite{blm+01}.  
Many GRBs have been localized by the Interplanetary Network (IPN) of spacecrafts,
whose synergy allows triangulation of the GRB position, yielding in many cases
arcminute-sized error areas, although with delays seldom shorter than 12 hours,
and often larger than 24 hours \cite{hbk+99a,hbk+99b}. For most cases, a
refinement of the localization error box has been possible after detection of the
X-ray afterglow by the BeppoSAX Narrow Field Instruments, by the RXTE PCA, or by
Chandra (see also chapter by F. Frontera).

Optical afterglows are generally identified for being previously unknown sources and for
their variability, either by comparing deep images of the GRB field, acquired soon
after the GRB detection, with the Digitized Sky Survey (DSS), or, for transients
fainter than the DSS limit, with images obtained at later epochs.  An alternative
technique for the selection of GRB counterpart candidates is based on the
characteristic power-law spectral shape of afterglows, and can be successfully
employed when sufficient color information is available \cite{jer01,gfh+01}. The
method may suffer from redshift-dependent biases and possible contamination by
other classes of sources; however, it can be advantageously applied to a single
set of images, instead of two (or more) sets of images taken at different epochs.

It should be specified that so far optical counterparts have been detected only
for long duration GRBs, which represent one of the two populations into which GRBs
are subdivided (see, however, \cite{mfj+98,brc01} for the possible existence of a
third class). According to their duration, the GRBs of the BATSE sample are
divided into long (75\% of the total) and short events (25\%), with average
durations of $\sim$20 s and $\sim$0.2 s, respectively. This bimodality is
reflected also in the spectral hardness, with long GRBs tending to be softer than
sub-second GRBs \cite{kmf+93}, and may be due to a different origin of the two
classes of sources. Sub-second GRBs could not be accurately localized by the
BeppoSAX WFC \cite{gsc+01}; four of them have been localized with arcminute
precision by the IPN, and followed up in the optical with delays no shorter than
$\sim$20 hours. No afterglow has been detected; the upper limits show that, within
the limited statistics, the optical afterglow behavior of sub-second GRBs may not
differ from that of long ones \cite{hbc+01,gah+01}.  The GRB000301C, which was
detected by the IPN with a 2 s duration and exhibited a bright, variable
counterpart (see \S 3.2), cannot be unambiguously classified as a long, sub-second
hard, or intermediate GRB \cite{jfg+01}.

Considering only GRBs for which the angular localization is better than $\sim$30
arcmin$^2$ and was disseminated within 24 hours, the statistics of detected
optical afterglows is $\sim$40\% of the total.  Therefore, many GRBs are optically
``dark", though nearly all of these have X-ray and/or radio afterglows. Many
causes can concur to make optical searches unsuccessful (see also \cite{dfk+01}).  
Optical afterglows can be intrinsically faint, rapidly decaying, or dim because of
the large redshift, causing the Ly$\alpha$ break to affect the optical spectrum.
Therefore, in some cases, the lack of an optical detection may be due to the
insufficient sensitivity of the search \cite{tbf+00,fjg+01,fkb+99}.  The different
decay rates of the optical afterglows on the one hand and the rather wide range of
magnitudes measured, at equal intervals after the GRB, for optical counterparts
with comparable decay rates on the other (see Table~1), indicate that the chance
of detection may be very different from case to case for similarly prompt and deep
exposures. This, together with the lack of a straightforward correlation between
optical emission and gamma-ray brightness of the prompt event, makes it difficult
to predict the detection level and to devise an optimal observing strategy.

In addition, optical afterglows may be affected by extinction in the plane of our
Galaxy (which is instead transparent to gamma- and hard X-rays and to radio
wavelengths), and by absorption in their host galaxies, which makes their
detection all the more challenging.  By simulating the absorption experienced at
the center of a dense dust clump, similar to those found in star forming regions
in our Galaxy and in external ones, in a number of directions randomly
distributed, Lamb \cite{dql01} determined that in only 35\% of the lines of sight the
optical
depth is $\tau = 1$, while in the remainder it is $\tau \gg 1$.  
This statistic is consistent with the percentage of dark GRBs, supporting the idea
that local dust absorption may hamper or completely prevent optical detection of
the GRB afterglow and strengthening the importance of infrared observations (see
\S 4.1).  The presence of substantial quantities of dust at the burst explosion
site favors, in turn, the association of GRBs with star forming regions (see \S
5). Although the result of this test would relate all dark GRBs to the effect of
dust extinction, perhaps this is only one of the possible causes for a failed
optical detection. An alternative view \cite{lcg01} is that dark GRBs can be
heavily extincted only if dust sublimation by the strong UV/optical \cite{ewbd00}
and X-ray radiation \cite{fkr01} following the explosion does not play a
significant role.  If dust destruction around the burst site is important
\cite{tgrw01}, then dark GRBs should belong to a distinct population with respect
to GRBs with detected optical afterglows.

\section{Temporal characteristics}

\subsection{The Optical ``Flash"}

Prompt emission at optical and near-infrared wavelengths simultaneous with a GRB,
or delayed by a few seconds, is expected to take place as a consequence of a
reverse shock propagating into the explosion ejecta, and is therefore distinct
from the afterglow, which is produced by the interaction of the forward shock with
the interstellar medium \cite{mrp94,pmmr97,appm98,rstp99a}.
This low energy early emission can be very bright, in principle up to
the $\sim$5th magnitude in the visual band, for an intense GRB at $z
\sim 1$, but
is
expected to
last only tens of seconds.

Since current instrumentation disseminates accurate GRB localizations with delays
of a few hours, GRB fields are usually imaged in optical starting no earlier than
some hours after the explosion.  Even if dissemination occurred in real time (as was the
case for
the large BATSE error boxes), the
observer reaction at a traditional optical telescope would still take at least
several minutes.  Small robotic telescopes, which can slew automatically and
rapidly to the GRB position in response to a localization delivered in real time,
are therefore the most efficient instruments for GRB follow-up at the earliest
epochs.                         

\begin{figure}
\begin{center} 
\includegraphics[width=0.6\textwidth]{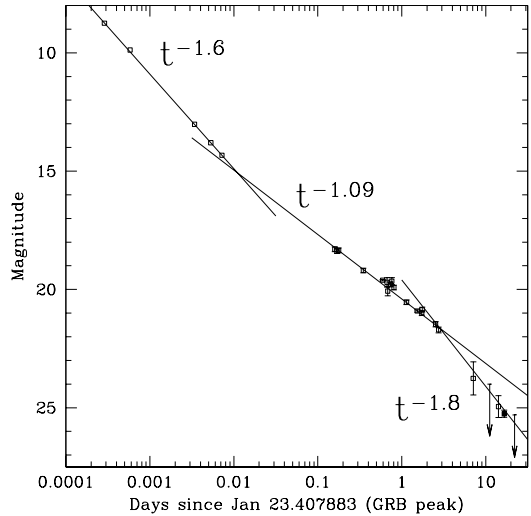}
\end{center} 

\caption[]{$R$-band light curve of the GRB990123 afterglow. All points, except for
the HST point (rightmost filled square), represent measurements taken from the
ground (see \cite{ftm+99} for references)  and are reduced to a common flux
standard with the galaxy flux subtracted. Error bars (1 $\sigma$) are shown where
available, and arrows indicate 95\% confidence upper limits (from \cite{ftm+99}).}

\label{lc990123} 
\end{figure}

On 23 January 1999, an optical flare was
detected by the ROTSE-I robotic telescope (consisting of a two-by-two array of 35
mm lenses), starting 22 seconds after the onset of a GRB which triggered both
BeppoSAX and BATSE. At maximum, the optical transient reached $V \simeq 9$
\cite{abb+99}, implying a power output in the optical of about $\sim$1\% of that
at the gamma-ray energies \cite{gbw+99}, in agreement with the reverse shock
interpretation, and allowing an estimate of the plasma initial Lorentz factor
\cite{rstp99b}.  In Fig.~\ref{lc990123} the ROTSE-I data are reported (from
immediately after the light maximum, to $\sim$10 minutes after the GRB, before the
decreasing flux level becomes undetectable), along with the measurements of the
successive afterglow taken at bigger telescopes. The ROTSE-I points are fitted by
a steeper temporal power-law than the afterglow points, indicating two different
radiation mechanisms. Owing possibly to the exceptional brightness of the event,
to the rapidity of the ROTSE-I slew, to the precise BeppoSAX localization (which
allowed identification of the transient in the $16^{\circ} \times 16^{\circ}$
ROTSE-I CCD image), and to the limited sensitivity of state-of-the-art robotic
telescopes, GRB990123 still remains the unique case of detection of an ``optical
flash" simultaneous with the GRB event itself.  Prompt searches of other GRB error
boxes both with ROTSE-I and with other robotic systems have yielded no detection
to limiting magnitudes spanning from $\sim$4 to $\sim$15 at epochs comprised
between 10 seconds and 30 minutes after the GRB
\cite{abb+00,kab+01,pwa+97,ppw+99,csg+01,bab+01}.

\subsection{Optical afterglow emission}

At epochs between a few hours and $\sim$1 day after the GRB the afterglow decays
following approximately a temporal power-law $t^{-\alpha}$ with an index $\alpha$
ranging from $\sim$0.7 to $\sim$2 (see Table~1, Col. 4, and Fig.\ref{lc990123}). This
behavior was predicted before the detection of the first afterglow as a consequence of
the simplest version of the fireball model \cite{pmmr97}.  According to this scenario
(see chapter by E. Waxman), the GRB explosion triggers a relativistic forward shock
which develops into the interstellar medium (or in a wind pre-ejected by the GRB
stellar progenitor \cite{tp99,rczl99}) and accelerates the particles. These interact
with the magnetic field and radiate at all frequencies through the synchrotron
mechanism (significant contribution from the inverse Compton scattering emission is
also predicted and observed in some cases \cite{rsae01,appk00,hys+01}). Linear
polarization, measured in the optical afterglows of GRB990510 and GRB990712 at the
level of a few percent \cite{wvg+99,clg+99,rvw+00}, represents a good test of the
synchrotron mechanism \cite{alrp98,agew99} and of collimated emission \cite{gg01} (see
below).
Near-infrared polarimetry of GRB000301C
yielded an upper limit of 30\%. Although not very constraining, this is consistent
with a synchrotron origin of the continuum in a relativistic jet \cite{sfk+01}.

On average, the initial luminosities of GRB optical afterglows are two orders of
magnitude larger than maximum supernova luminosities, and obviously outshine their
parent galaxies.  At longer intervals after the GRB, when the flux of the transient
subsides under the brightness of the host galaxy, it is possible to measure the
magnitude of the latter. In some cases (notably GRB970228, GRB980326 and GRB011121),
the light curve of the optical afterglow exhibits a rebrightening with respect to a
power-law behavior at few weeks after the GRB.  Following claims for the possible
association of GRB980425 with the close-by ($z = 0.0085$)  SN1998bw
\cite{gvv+98,kfw+98} (see also chapter by T. Galama), the light curves of those
optical transients have been decomposed into a non-thermal, pure afterglow
contribution and a supernova profile, using SN1998bw as a template, appropriately
redshifted. In some cases the results are convincing
\cite{bkd+99,gtv+00,der99,csg+01,bkp+02}, while in other cases they are not decisive
\cite{svb+00,bhj+01,jhb+01}. Systematic decomposition into non-thermal and supernova
emission components has been attempted also by Dado et al. \cite{ddd01,ddd02a,ddd02b}
for the afterglows of GRBs with known redshift, in the framework of the Cannonball
model, and good results have been obtained in most cases. The afterglow magnitudes
reported in Table~1 (Col. 6) have been obtained from a fit with a single or double
power-law after subtraction of the host galaxy flux, and in some cases after
decomposition of a possibly underlying supernova (see references for individual cases
in Col. 9).  In those cases where the few optical measurements did not allow a proper
fit, the optical magnitude measured closest to 1 day after explosion was used (e.g.,
GRB980613).

In a large fraction of the best monitored optical and/or near-infrared afterglows
the initial power-law decline steepens at times ranging from $\sim$0.5 to $\sim$5
days after the GRB explosion. The effect is clearly seen as a smooth increase of
the flux decay rate
\cite{czc+99,ftm+99,kdo+99,sgk+99,hbf+99,imc+99,hum+00,mbb+00,jraf99,jfg+01,sbrc00,fgd+01,hys+01,phg+01,sgj+01,mpp+01},
and is suggested also by the X-ray data in a few cases
\cite{psa+01,hys+01,pgg+01,zka+01}. In Table~1 (Cols. 4 and 5)  the early and late
temporal indices are reported as determined via empirical fits to the optical
light curves with double power-laws (see also Fig.\ref{lc990123}).  The change in
the temporal slope is thought to witness the presence of a decelerating jet.
Collimation of the radiation in a jet structure would reduce the huge energy
outputs ($\sim 10^{52}-10^{54}$ erg) derived from the observed gamma-ray
brightnesses and the measured distances of GRBs, in the assumption of isotropy,
and thus help resolving the paradox of energy conversion efficiency
\cite{nsad95,pm99,tp99,fks+01}. When the aperture of the radiation cone (beaming
angle),
which progressively increases as the relativistic plasma decelerates, becomes
larger than the jet opening angle, the observer perceives a faster light dimming,
independent of wavelength, due to the jet edge becoming visible and/or to jet
sideways expansion \cite{pmmr99,sph99,jer99}.  The change in fading rate is
however smooth, due to light travel time effects at the ending surface of the jet
\cite{appm99,msb00}.  The steepening of the afterglow light curve would then be a
probe of the GRB and afterglow emitting geometry.  Stanek et al. \cite{sgj+01}
note an anti-correlation between the slope change $\Delta\alpha$ and the isotropic
gamma-ray energy of the burst, suggesting that the different jet opening angle may
be responsible for it.  Specific jet models for individual cases have been
proposed \cite{ap01,bdf+01,bsf+00}, and afterglow emission from jets has
been modeled in the firecone scenario as a 
function of
the viewing angle \cite{rlr02,bzpm02,gpkw02}, as also suggested by Dado et al.
\cite{ddd01}.

If jets are unavoidable to relax the energy crisis in GRBs and a light curve steepening
is their signature, one may wonder why all observed optical afterglows do not exhibit a
detectable steepening in their light curve.  This may simply be due to undersampling:  
when not detected, the steepening may have occurred at early, not well-sampled epochs
(many afterglows are described by power-laws with temporal indices steeper than 2, see
Table~1), or at late epochs, when the afterglow behavior is significantly contaminated
by the emerging host galaxy or possible supernova, so that discerning a decay rate
variation is more difficult (see e.g., \cite{fpg+00}). 
On the other hand, light curve steepening cannot be univocally ascribed to a
decelerating jet, but may be caused instead, or in addition, by the transition of
a spectral break through the observing frequency band \cite{spn98} (see \S 4.1) or
by the propagation of the external shock in a non homogeneous medium
\cite{pmr98,mrw98,rczl99,jhb+01} (although in these cases the steepening would be
frequency dependent; see however \cite{pkap00} for detectability of a jet in a
stratified medium), or by the transition of the plasma kinematic conditions from
relativistic to Newtonian in a dense medium \cite{zdtl99,zdtl01,mpp+01,hdl00}.  
In some cases the interpretation is not unique \cite{hkpb99,zlrc01}, although
simultaneous multiwavelength observations may resolve the ambiguity
\cite{rczl00,hys+01,appk01a,pgg+01}. 
We finally note that some months after the GRB a flattening of the afterglow 
light curve may be expected instead \cite{mlew00}.

Optical interday or intraday variations superimposed to the overall afterglow
decline are rarely detected, because of the limited photometric precision of
the
measurements. Two cases where significant deviations from a steady decay have been
observed during the optical monitoring are GRB970508, which exhibited an initial
shallow decline, followed by a 2-day re-bursting of a factor $\sim$5 amplitude
\cite{cgb+98,pjg+98,ggv+98} correlated with an X-ray flare \cite{paa+98}, and
GRB000301C, which showed intraday achromatic variability of 20-30\% amplitude
\cite{bsf+00,mbb+00}.  For both events, an interpretation based on microlensing
has been proposed \cite{ddd01,gls00,ggl01}. Rapid variability can otherwise be
produced by small scale inhomogeneities of the plasma flow, or irregularities of
the external medium in which the blast wave propagates \cite{xwal00,hsg+02}, or local
re-energization
episodes \cite{pmr98}.

\section{Spectral properties}

\subsection{The infrared-optical afterglow continuum}

The classical fireball model has specific predictions for the
temporal evolution of the broad-band spectral shape \cite{spn98}. This has been
modified to include the detectable effects of the presence of a jet \cite{sph99}.  
Both the spectral slope and the temporal decay rate depend on the index $p$ of the
electron energy distribution $N(\gamma) \propto \gamma^{-p}$ (where $\gamma$
denotes the electron energy above a certain cutoff).  The radio-to-X-ray spectral
shape is characterized by smooth breaks at typical frequencies (self-absorption,
peak and cooling frequencies), which evolve with time in a predicted way
\cite{jgrs01}, so that simultaneous multiwavelength observations at various epochs
during the evolution of the afterglow allow the measurement of the spectral slopes
and breaks and the estimate of the relevant physical parameters of the afterglow
(see e.g., \cite{rwtg99,appk01a,appk01b}, and chapter by F. Frontera).

When the optical photometric observations are accurate and sufficiently
extended in time to make a good signal-to-noise ratio measurement of the spectral
and temporal slopes possible, they show that the spectra of some afterglows, corrected
for Galactic extinction, are steeper (i.e., redder) than expected from the fireball 
theory
based on comparison with the temporal decay rate.  This has been commonly
attributed to absorption intrinsic to the source or, especially for GRBs at very
high redshift, intervening along the line of sight \cite{rkf+98,der01a,shpm01}.
A small amount of reddening
by dust in the GRB host galaxy has been invoked in many cases to reconcile the
observations with the theoretical scenarios, using extinction curves typical of our
own Galaxy, of star-forming galaxies, or of the LMC and SMC
\cite{ppm+98,vgo+99,daa+00,mpp+01,hys+01,ltv+01,jfg+01,ddd02a,ddd02b}.  While even a
moderate
quantity of dust at the GRB source redshift may significantly attenuate the
observed optical spectrum (which corresponds, at the average $z \sim 1$, to
rest-frame ultraviolet wavelengths), or even completely obscure it (see \S 2),
near-infrared data are less affected and may be more effective in determining the
overall afterglow spectrum, when combined with data at other frequencies
\cite{ppm+98,daa+00,lcr99}. Observations in the near-infrared range are therefore
critical for the study of afterglows.

\subsection{Absorption features}

\begin{figure}[]
\begin{center}
\includegraphics[width=0.8\textwidth]{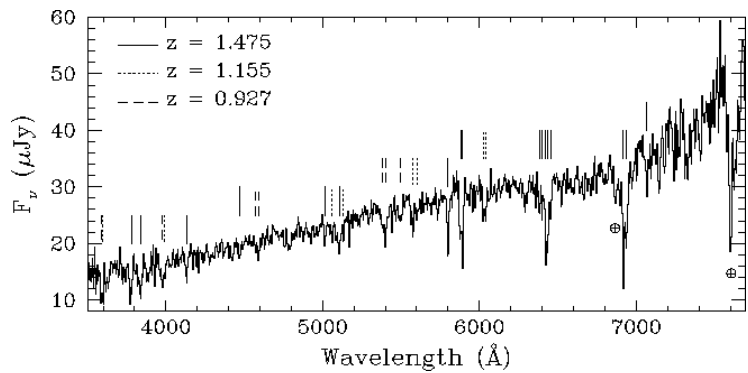}
\end{center}
\caption[]{Low resolution spectrum of the GRB010222 afterglow taken at the Telescopio
Nazionale Galileo, corrected for Galactic extinction.
Absorption lines from systems at three different redshifts are detected.
Telluric features are marked with the symbol $\oplus$ (from \cite{mpp+01}).}
\label{010222spec}
\end{figure}

Low and medium resolution spectra of bright optical afterglows have allowed, in a
number of cases, the detection of absorption lines of metallic species caused by
intervening absorbers, and the measurement of lower limits to the GRB redshift
(see Table~1 and Fig.\ref{010222spec}). Frequently, the evidence of a
low-ionization, high density medium (e.g., related to the detection of Mg~I in
absorption) suggests that the absorbing system is actually the host galaxy
\cite{mdk+97,vfk+01,mpp+01}. Spectroscopy of the likely host galaxy generally
shows that the highest absorption redshift coincides with the redshift of the
galaxy emission lines, confirming the association of the GRB with the proposed
host. No variability of the absorption line equivalent widths is detected at the
20\% level (which represents the 1-$\sigma$ uncertainty on the measurements) in
time scales of some hours to few days \cite{vfk+01}.

\begin{figure}[]
\begin{center}
\includegraphics[width=0.8\textwidth]{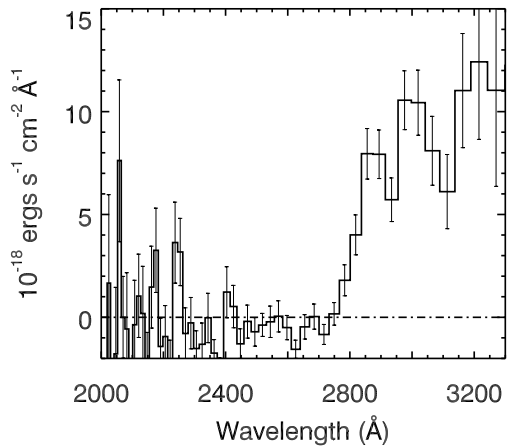}
\end{center}

\caption[]{Deconvolved, flux-calibrated, ultraviolet spectrum of GRB 000301C taken
with the HST STIS NUV MAMA prism. A break is clearly seen at 2797 \AA. If caused
by the onset of Lyman continuum absorption due to H I gas associated with the host
galaxy, the redshift is $z = 2.067 \pm 0.025$. This matches with the ground-based
measurement of the redshift (from \cite{sfg+01}).}

\label{smette}
\end{figure}

\begin{figure}[]
\begin{center}
\includegraphics[width=0.7\textwidth]{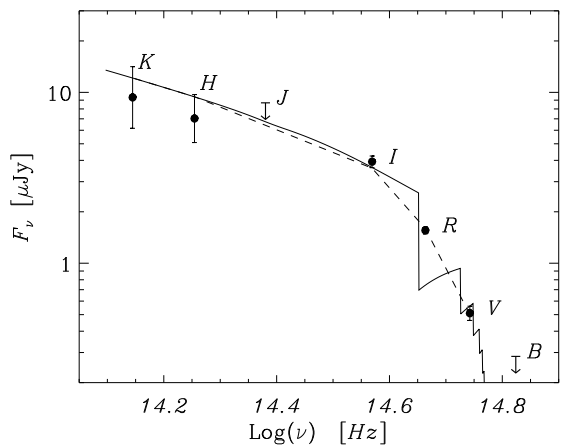}
\end{center}

\caption[]{Spectral energy distribution of the GRB000131 afterglow, corrected for
Galactic reddening, as derived from broad band VLT and NTT photometry. The
uncertainties of the $H$- and $K$-band fluxes include the formal error from the
extrapolation of the light curves back to the epoch of the optical measurements,
$t = 3.5$ days. A fit with a power-law spectrum with Ly$\alpha$ forest absorption
and SMC reddening is shown as a dashed curve. This yields $A_V = 0.18$, when an
intrinsic spectral slope $\beta = 0.70$ and a redshift of 4.5 is assumed. The
solid curve shows the corresponding spectrum with its Lyman absorption edges (from
\cite{ajp+00}).}

\label{000131spec}
\end{figure}

For GRB000301C and GRB000131 the redshift has been determined through
identification of absorption edges with the Lyman limit and intervening Ly$\alpha$
forest, respectively. The afterglow of the former GRB is the only one which has
been observed by HST at ultraviolet wavelengths: in the low-resolution spectrum
taken 5 days after explosion with the STIS instrument equipped with the NUV MAMA
prism a discontinuity at $\sim$2800 \AA\ is clearly detected, which has been
identified with the hydrogen ionization edge \cite{sfg+01}, implying $z \simeq 2$
(see Fig.\ref{smette}). The measurement was then confirmed and refined from
optical spectroscopy at the Keck and VLT telescopes \cite{cdd+00,jfg+01}.  The
redshift of GRB000131, $z \simeq 4.5$, was determined photometrically from
simultaneous near-infrared and optical observations (see Fig.\ref{000131spec}),
and supported by optical spectrophotometry \cite{ajp+00}.

The distance of GRB980329 is controversial: the afterglow photometry suggests a
redshift as large as $z \sim 5$ or lower, $3 \simlt z \simlt 4.4$, according to
whether the observed continuum suppression shortward of $\sim$6500 \AA\ is
identified with Ly$\alpha$ intervening absorption \cite{asf99}, or with molecular
hydrogen dissociation by the strong initial ultraviolet flash \cite{btd00}. A
redshift $z < 3.9$ would be suggested by the absence of the Ly$\alpha$ break in
the host galaxy spectrum \cite{dql99,dkb+01}.

The redshifts measured so far either with spectroscopy or broad-band photometry
span the range $\sim$0.4 to $\sim$4 (Table~1), excluding the peculiar case of GRB980425
(see
chapters by T. Galama and K. Iwamoto), and prove that long duration GRBs have an
extragalactic, cosmological origin, which makes their early bright optical
afterglows excellent probes of the high redshift universe.

\section{Host galaxies}

For almost every well studied optical afterglow, deep late epoch optical
or near-infrared observations from the ground or with HST have detected a
galaxy close to the point-like optical transient or at its position after
it has faded away (see Table~1).  Host galaxies have been detected also
for some dark GRBs with arcsecond afterglow localizations from radio
telescopes or from Chandra.  Few observations of host galaxies have been
made at longer wavelengths \cite{stv+99,hlm+00,bkf+01,fbm+01}). The hosts
optical magnitudes (and upper limits) are consistent with those expected
for a reasonable redshift distribution and galaxy population
\cite{dhaf99}. This solves the ``no-host galaxy" problem
\cite{bes92,dbdh98,bhs99,sbl97}, which, a posteriori, turns out to be
clearly related to the very faint flux of the host galaxies, which are
usually detected only with long exposures at telescopes larger than 2m.

\begin{figure}[]
\begin{center}
\includegraphics[width=0.5\textwidth]{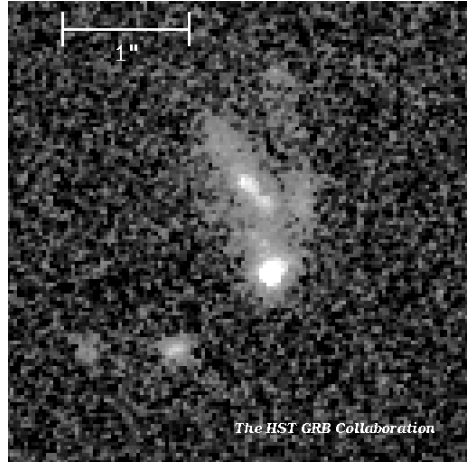}
\end{center}

\caption[]{HST STIS image of the host galaxy of GRB990123 in white light. North is
up, and east is to the left. The galaxy morphology is irregular. The optical
transient is the bright point-like source on the southern edge of the galaxy (from
\cite{ftm+99}).}

\label{990123host}
\end{figure}

HST observations taken at early stages of the afterglow evolution, when
the transient is still bright (see Fig.\ref{990123host}), show that this
lies always within the stellar field of its host galaxy. If only late
epoch HST images are available, their comparison with accurate astrometry
of the bright transient on early epoch ground-based images still yields
projected angular offsets of a fraction of an arcsecond between the
transient and the galaxy center. When normalized to the individual hosts
half-light radii, the median offset is 0.98 \cite{bkd01}.  This has led to
the conclusion that GRBs are associated with star forming regions,
which would
support their origin as hypernovae \cite{bp98} (or collapsars
\cite{amsw99}) or as {\it supranovae} \cite{mvls98}, as opposed to
progenitor scenarios which envisage the explosion as taking place at many
kiloparsecs from the parent galaxy, in its halo (like binary neutron star
systems, see chapter by E. Waxman).

\begin{figure}[]
\begin{center}
\includegraphics[width=0.8\textwidth]{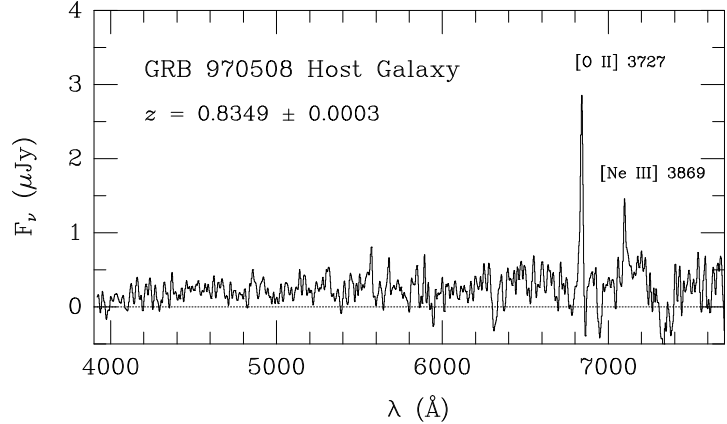}
\end{center}

\caption[]{Spectrum of the host galaxy of GRB970508, obtained at the Keck telescope.  
Prominent emission lines are labeled (from \cite{bdkf98}).}

\label{970508host}
\end{figure}

There is specific evidence that host galaxies of GRBs undergo strong star
formation: 1) their integrated colors are remarkably blue \cite{fpt+99,ftm+99}; 2)
they are usually underluminous (luminosities of a fraction of the characteristic
luminosity $L_*$ of the Schechter \cite{ps76} luminosity function), small, and
have often a compact morphology \cite{hys+01,pfb+98,fpg+00}, which are
characteristics common to galaxies hosting star formation at the typical GRB
redshifts, $z \sim 1$ \cite{abmjr92,ggk+97}; 3) their spectra exhibit star
formation emission lines like [O II], [Ne III] (Fig.\ref{970508host}), [O
III], Ly$\alpha$ and Balmer series \cite{bdkf98,kdr+98,dkb+98,vfk+01,fmt+02}; 4) their
star
formation rates, derived
either from the emission line intensities or from the ultraviolet rest frame
galaxy continuum are high, compared with the galaxy size
\cite{fpt+99,dfk+01,dkb+98,hfh+01,vfk+01,dkb+01,sh01}, although sometimes
substantially obscured \cite{dfk+01,dkb+01} and possibly measurable only at rest
frame far-infrared wavelengths \cite{fbm+01}; 5) they sometimes have morphologies
consistent with being mergers or interacting systems \cite{ftm+99,dfk+01,dbk01},
where star formation is enhanced.  Furthermore, the imprints of local extinction
on the afterglow spectra (see \S 4.1) point to dusty and likely star forming
environments as the favored GRB explosion sites. Finally, the observed host galaxy
magnitudes and redshifts are consistent with a model in which the comoving rate
density of GRBs is proportional to the cosmic star formation rate density
\cite{dhaf99,smhm98,kth98,rft01,wglr01}. These suggestions are consistent with the
fact that most measured GRB redshifts are around $z \sim 1$, where the cosmic star
formation rate is one order of magnitude larger than locally \cite{dql00}. Since
this is predicted to increase monotonically back to $z \sim 10$, one may expect
that, given the opportunity of detecting GRBs up to that redshift, it would be
possible to select the youngest star forming galaxies in the universe.

\section{Conclusions: open problems and future prospects}

The enormous progress achieved over the last few years through the optical
afterglow follow-up study has also indicated some fundamental, unclear aspects of
GRB and afterglow physics: 1) the conversion of energy into radiation; 2) the
structure and geometry of the emitting regions; 3) the nature, density and
composition of the circumburst medium; 4) the cosmological evolution of the GRB
population.  The solution of these problems would ultimately lead to unveiling the
major unknown, i.e. the identity of GRB progenitors. While at the present stage
these issues remain a matter of investigation, current knowledge suggests the
observational approach to tackle them most effectively.

The most serious limitation of present day observations is the substantial
temporal delay between GRB trigger and follow-up of its field at lower frequencies
(a few hours). This causes a lack of sampling of the initial portions of the light
curves. Bridging this gap will be possible when real-time disseminated
localizations, made available by the coming generation of high energy satellites,
can be rapidly followed by small, fast reacting telescopes with suitably large
fields of view.  Thanks to the UVOT camera onboard the GRB mission SWIFT (to be
launched in 2003) and to the advent of ground-based robotic telescopes
\cite{csb+99,ppw+99,zcr+01,epep01,bab+01}, the monitoring will start as
early as few tens of seconds after GRB detection, allowing astronomers to catch
the transient counterpart in its maximum emission state, and to follow the
temporal evolution of its optical-to-infrared continuum.  It is at these early
epochs that the models differ most in their predictions
\cite{ddd01,appk00,rstp99a,spn98,pmmr99}, and the strength of the signal can
discriminate them with the highest confidence.  Early, simultaneous
optical/near-infrared searches will either detect more counterparts, and reduce
the number of dark GRBs, or put stronger constraints on the ``darkness".  
With respect to robotic systems, all-sky
optical monitoring cameras \cite{hpma94,bp01} 
present the advantage that they would be independent from spacecraft
triggers, and therefore they could detect possible precursors of GRBs and
``burstless" afterglows\footnote{Highly collimated jets misaligned with respect to
the line of sight would prevent detection of the GRB, while its optical afterglow
may become detectable after jet spreading \cite{mrw98}.}, which would be obviously
missed by robotic telescopes.

Weeks after the GRB explosion, the light curve will result from the sum of different
components of comparable brightness: the fading afterglow, the possible supernova rising
to maximum light, and the host galaxy. To disentangle these contributions we will need
sensitive and densely sampled photometric observations at late epochs in the optical and
infrared. These will also possibly discern the presence of a light echo, due to dust
scattering \cite{aerb00,der01b} or sublimation \cite{ewbd00,bvab01}.  This task is a
prerogative of the sensitive, high resolution optical and near-infrared cameras in
space, such as the HST WFPC2, STIS, NICMOS, and the newly-deployed Advanced Camera for
Surveys.  The future NGST will collect the legacy of these instruments and push the
research on GRB late afterglows and hosts toward even larger redshifts.

The measurement of a large number of redshifts, through early absorption spectra
of optical afterglows as well as emission line spectra of the host galaxies, is
necessary to construct a luminosity function of GRBs to be compared with models of
star forming rate evolution. This will allow us to test the link between GRBs and
star formation history up to very high redshifts, many hints of which have been so
far collected \cite{dql00}.  A primary role in early and late sensitive optical
and infrared spectroscopy will be played by the ground-based 4 to 8 m class
telescopes in both hemispheres. Early bright counterparts will be excellent
targets for spectroscopic monitoring: variations in the equivalent widths of the
absorption lines will be measured with good signal-to-noise ratio, thus making it
possible to place constraints on the density and distribution of the circumburst
medium \cite{lpg01,rpal98,btdlh01}, which is a critical diagnostic of the GRB
progenitor. Furthermore, the intense initial optical flares associated with GRBs
will also act as background ``light bulbs" to probe the ionization state and
metallicity of the intergalactic medium through high resolution spectroscopy. More
generally, multiwavelength observations of GRBs allow a series of cosmological
tests on a wide range of redshifts \cite{dldr00,dldr01,rpaa00} (see also chapter
by A. Loeb).

Prompt, long and intensive polarimetric monitoring of afterglows will detect
possible changes of polarized light percentage and position angle and thereby set
constraints on the most important open issue of afterglow physics, the generation
of magnetic fields \cite{agew99,ag99,ag01,mmal99,ggdl99}. Moreover, knowledge of
the magnetic field geometry and of the circumburst density profile will be
instrumental to defining the structure of the jet and its interaction with the
ambient medium.

Finally, real time dissemination of accurate localizations of sub-second GRBs and
the prompt follow-up of these fields in the optical will hopefully afford
detection of their elusive counterparts, and will allow us to get clear insights
into their genesis and physics.

\section{Acknowledgments}

I am grateful to the members of the GRACE Consortium led by E. van den Heuvel for
a
longstanding, fruitful collaboration, and in particular to F. Frontera, A.
Fruchter, N. Masetti and E. Palazzi, who have been my closest collaborators in GRB
optical counterparts research since the beginning of the ``afterglow era". I thank
S. Recchi for valuable inputs, and A. De R\'ujula, A. Fruchter, P. Mazzali and E.
Waxman for a critical reading of the manuscript.  I would like to dedicate this
review to the memory of Jan van Paradijs.

\begin{table}

{\small
\caption{Parameters of GRB Optical Counterparts$^a$}
\begin{center}
\renewcommand{\arraystretch}{1.4}
\setlength\tabcolsep{2.3pt}
\begin{tabular}{ccccccccc}
\hline\noalign{\smallskip}
GRB & Instr.$^b$ & $z$& $\alpha_1^c$ & $\alpha_2^c$ & $R_{OT}^{d,e}$ 
& $R_{host}^{d,f}$ & $A_R^g$ & Refs.\\  
& & & &  & (mag) & (mag) & (mag) & \\ 
\noalign{\smallskip}
\hline
\noalign{\smallskip}
\\
\hline       
970228 & BeppoSAX & 0.695 & 1.7 && $20.3 \pm 0.2$ & $24.6 \pm 0.2$ & 0.63 &
\cite{bdk01,gtv+00,fpt+99} \\ 
970508 & BeppoSAX & 0.835 & 1.3 && $21.0^h \pm 0.1$ & $25.0 \pm 0.2$ & 0.13
& \cite{mdk+97,bdkf98,fpg+00} \\
970828 & BATSE/RXTE & 0.9578 & ... & ... &  $>23.7$ (4 h) & $25.1 \pm 0.3$ & 0.10 &
\cite{dfk+01,ggv+98a,bkg+01} \\ 
971214 & BeppoSAX & 3.418 & 1.4 && $23.0 \pm 0.1$ & $26.2 \pm 0.2$ & 0.04 & 
\cite{kdr+98,hthc98,odk+98} \\ 
980326 & BeppoSAX & $\sim 1^i$? & 2.0 && $22.8 \pm 0.1$ & $V = 29.0 \pm 0.3$ 
& 0.21 & \cite{bkd+99,ggv+98b,fvn01}\\  
980329 & BeppoSAX & 3-5 & 1.2 && $23.7 \pm 0.2$ & $27.8 \pm 0.3$ & 0.19 &
\cite{asf99,lcr99,ppm+98,rlm99,hth+00a} \\
980425 & BeppoSAX & 0.0085 & ... & ... & $15.60 \pm 0.05$ & $14.11 \pm 0.05$ & 0.17 & 
\cite{tscg98,gvv+98,fha+00} \\
980519 & BATSE/WFC & ... & 1.73 & 2.22  & $20.64 \pm 0.03$ &
$\sim 25.5$ & 0.69 & \cite{jhb+01,hpja99,hft+00a} \\   
980613 & BeppoSAX & 1.097 & $\sim 0.2$ & $\sim 1$ & $23.1 \pm 0.1$ & $24.3 \pm
0.2$ & 0.23 & \cite{dbk01,hapj98,dkoe98,hf98,sh01} \\ 
980703 & BATSE/RXTE & 0.966 & 1.4 && $21.00 \pm 0.05$ & $22.4 \pm 0.1$ & 0.15 &
\cite{dkb+98,czg+99,hfh+01} \\
981226 & BeppoSAX & ... & ... & ... & $>23$ (10 h) & $24.2 \pm 0.1$ & 0.06 &
\cite{lhp+99,hta+00a}\\
990123 & BeppoSAX & $>$1.6004 & 1.13 & 1.8 & $20.4 \pm 0.1$ & $23.9 \pm 0.1$
& 0.04 & \cite{kdo+99,czc+99,ftm+99}\\
990308 & BATSE/RXTE & ... & $\sim$1.2 & & $20.7 \pm 0.1$& $>$28.4 & 0.07 &
\cite{ssh+99,hft+00b}\\         
990506 & BATSE/RXTE & 1.3 & ... & ... & $>19$ (1 h) & $24.8 \pm 0.3$ & 0.18 &
\cite{bfs01,zz99,hta+00b}\\        
990510 & BATSE/WFC & $>$1.619 & 0.82 & 2.18 & $19.00 \pm 0.05$ & $V =
27.4 \pm 0.3$ & 0.53 & \cite{vfk+01,hbf+99,fhp00} \\
990705 & BeppoSAX & $\sim$0.86 & 1.7 & $>$2.6 & $H \sim 19$ & $22.0 \pm 0.1$ & 0.20 &
\cite{afv+00,mpp+00,scc+01}\\
990712 & BeppoSAX & 0.4331 & 0.97 &  & $21.25 \pm 0.05$ & $21.90 \pm 0.15$  &
0.08 & \cite{vfk+01,svb+00,fvhp00}\\
991208 & IPN & 0.706 & 2.3$^j$ & 3.2  & $18.5 \pm 0.1$ (2 d) & $24.2 \pm 0.2$ & 0.04 &
\cite{csg+01,fvsc00} \\ 
991216 & BATSE/RXTE & 1.02 & 1.0 & 1.8 & $18.0 \pm 0.1$ & $25.3 \pm 0.2$ & 1.64 &
\cite{vrh+99,pgg+00,hum+00,vffk00} \\
000131 & IPN & 4.5 & 2.3 & & $23.0 \pm 0.1$ (3 d) & $>$25.6 & 0.14 &
\cite{ajp+00} \\
000210 & BeppoSAX/CXO & 0.846 & ... & ... & $>22$ (12.4 hr) & $23.5 \pm 0.1$
& 0.05 & \cite{pfg+02} \\
000301C& IPN & 2.04 & 1.1 & 2.9 & $20.1 \pm 0.1$ (2 d) & $28.0\pm 0.3$ & 0.13 &
\cite{jfg+01,sfg+01,gls00,mbb+00,afpv01} \\
000418 & IPN & 1.118 & 1.22 & & $21.9 \pm 0.1$ (3 d) & $23.8 \pm 0.2$ & 0.08
& \cite{bdd+00,ksm+00,mfm+00} \\
000630 & IPN & ... & 1.0 & & $23.2 \pm 0.2$ & $26.7 \pm 0.2$ & 0.03 &
\cite{fjg+01,kekb01} \\   
000911 & IPN & 1.058 & 1.4 &  & $20.40 \pm 0.08$ (1.4 d) & $\sim$25 & 0.31 &
\cite{dkb+01,lcg+01,ppm+01} \\  
000926 & IPN & 2.037 & 1.5 & 2.3 & $19.50 \pm 0.02$ & $\sim$25
& 0.06 & \cite{cdk+00,fgd+01,phg+01,hgb+01} \\    
001007 & IPN & ... & 2.05 & & 20.2 (3.5 d) & $24.73 \pm 0.15$ & 0.11 &
\cite{pas00,ccg+02} \\
001011 & BeppoSAX & ... & 1.33 & & $22.4 \pm 0.1$ & $25.1 \pm 0.3$ & 0.26 & \cite{gfh+01}
\\  
001018 & IPN & ... & ... & ... & $>$22.6 ($>$2 d) & $24.50 \pm 0.09$ & 0.06 &
\cite{bdk+01,bdh+01} \\
010222 & BeppoSAX & 1.476 & 0.65 & 1.7 & $20.15 \pm 0.05$ &
$25.7 \pm 0.2$ & 0.06 & \cite{jpg+01,mpp+01,sgj+01,fbrl01} \\    
\hline
\end{tabular}
\end{center}
}
\end{table}
\pagebreak
\newpage
\begin{table}
{\small
\begin{center}
\renewcommand{\arraystretch}{1.4}
\setlength\tabcolsep{2.3pt}
\begin{tabular}{ccccccccc}
\multicolumn{9}{c}{{\bf Table 1}. (Continued)} \\
\noalign{\bigskip}
\hline\noalign{\smallskip}
GRB & Instr.$^b$ & $z$& $\alpha_1^c$ & $\alpha_2^c$ & $R_{OT}^{d,e}$ 
& $R_{host}^{d,f}$ & $A_R^g$ & Refs.\\  
& & & &  & (mag) & (mag) & (mag) & \\ 
\noalign{\smallskip}
\hline
\noalign{\smallskip}
\\
\hline       
010921 & \ HETE-2 \ & \ 0.45 \ & \ 1.59 \ & & \ $19.29 \pm 0.06$ \ &
\ $21.40 \pm 0.05$ \ & \ 0.38 \ & \cite{pkb+02} \\
011030$^k$ & \ BeppoSAX \ & \ ... \ & \ ... \ & \ ... \ & \ $>$21 (8 hr) \ 
& \ $V \sim 25$ \  & \ 1.00 \ & \cite{fpk+02,mppj+01} \\
011121 & \ BeppoSAX \ & \ 0.36 \ & \ 1.7 \ & & \ $19.47 \pm 0.05$ \ &
\ $24.70 \pm 0.05$ \ & \ 1.26 \ & \cite{igsw01,bkp+02} \\
011211 & \ BeppoSAX \ & \ 2.14 \ & \ 0.83 \ & \ 1.7 \ & \ $20.9 \pm 0.1$ \ 
& \ $25.0 \pm 0.3$ \ & \ 0.11 \ & \cite{fvrb01,hsg+02,brfh01}\\

\hline\noalign{\smallskip}
\multicolumn{9}{l}{$^a$ Detected before the end of year 2001.}\\
\multicolumn{9}{l}{$^b$ Instrument, or suite of instruments, which localized the
GRB.}\\
\multicolumn{9}{l}{$^c$ Temporal decay index in the $R$ band,  $f(t) \propto
t^{-\alpha}$.}\\
\multicolumn{9}{l}{$^d$ Magnitudes are in the Cousins system.}\\
\multicolumn{9}{l}{$^e$ Magnitude of the transient 1 day after the explosion (unless
noted
otherwise),}\\
\multicolumn{9}{l}{ ~~ corrected for the host galaxy contribution and for Galactic extinction.}\\
\multicolumn{9}{l}{$^f$ Host galaxy magnitude, corrected for Galactic extinction.}\\
\multicolumn{9}{l}{$^g$ Galactic extinction derived from the dust maps of Schlegel et
al. \cite{sfd98}, except for}\\
\multicolumn{9}{l}{ ~~ GRB970228 \cite{fpt+99}).}\\
\multicolumn{9}{l}{$^h$ At 1 day after the explosion, the afterglow was still rising,
therefore this}\\
\multicolumn{9}{l}{ ~~  magnitude is observed, and not derived from a fit.}\\
\multicolumn{9}{l}{$^i$ Estimate based on decomposition of the optical transient light
curve \cite{bkd+99}.}\\
\multicolumn{9}{l}{$^j$ Castro-Tirado et al. \cite{csg+01} argue that the slope may be 
flatter at times earlier than}\\
\multicolumn{9}{l}{ ~~ 2 days, based on an upper limit
obtained soon after the GRB by sky patrol films.}\\
\multicolumn{9}{l}{$^k$ This event belongs to the class of X-ray rich
GRBs (a.k.a. X-ray flashes), see}\\
\multicolumn{9}{l}{ ~~ chapter by F. Frontera.}\\

\end{tabular}
\end{center}

}
\label{Tab1.1a}
\end{table}


%

\end{document}